\documentclass[%
 aip,
 amsmath,amssymb,
 reprint,%
]{revtex4-2}

\usepackage{graphicx}
\usepackage{dcolumn}
\usepackage{bm}
\usepackage[utf8]{inputenc}
\usepackage[T2A,T1]{fontenc}
\usepackage[english]{babel}
\usepackage{mathptmx}
\usepackage{etoolbox}
\usepackage{cmap}
\usepackage{multirow}
\usepackage{array}
\usepackage{ragged2e}
\usepackage{tabularx}
\usepackage{hyperref}

\hypersetup{
    colorlinks=true,
    linkcolor=blue,
    citecolor=blue,
    filecolor=magenta,
    urlcolor=blue
}

\makeatletter
\def\@email#1#2{%
 \endgroup
 \patchcmd{\titleblock@produce}
  {\frontmatter@RRAPformat}
  {\frontmatter@RRAPformat{\produce@RRAP{*#1\href{mailto:#2}{#2}}}\frontmatter@RRAPformat}
  {}{}
}%
\makeatother

\begin{document}


\title[Electron energy gain in a DLA as a function of the base angle of a triangular grating structure]{Electron energy gain in a dielectric laser accelerator as a function of the base angle of a triangular grating structure}

\author{O.O. Svystunov}
\affiliation{%
National Science Center "Kharkiv Institute of Physics and Technology", Kharkiv, Ukraine
}%

\author{A.V. Vasyliev}
\affiliation{%
National Science Center "Kharkiv Institute of Physics and Technology", Kharkiv, Ukraine
}%

\author{I.V. Beznosenko}
\affiliation{%
National Science Center "Kharkiv Institute of Physics and Technology", Kharkiv, Ukraine
}%

\author{R.A. Melnichuk}
\affiliation{%
National Science Center "Kharkiv Institute of Physics and Technology", Kharkiv, Ukraine
}%

\author{G.V. Sotnikov}
\email{sotnikov@kipt.kharkov.ua}
\affiliation{%
National Science Center "Kharkiv Institute of Physics and Technology", Kharkiv, Ukraine
}%

\date{\today}

\begin{abstract}
Dielectric laser accelerators (DLAs) represent a compact and cost-effective alternative to conventional RF accelerators. Despite the various grating geometries already studied, a comprehensive investigation of triangular profiles, particularly the effect of the base angle of the saw-tooth grating profile, remains insufficient. This paper presents the results of numerical particle-in-cell (PIC) simulations of electron acceleration in DLAs based on double gratings with a triangular profile. One of the gratings, onto which the laser beam is incident, is transparent, while the second grating was either transparent or reflective for the laser pulse. The geometry of the first grating was fixed. A systematic study was conducted on the influence of the base angle of the second grating ($\alpha = 5^\circ - 44^\circ$), its spatial orientation, the presence of a reflective gold coating, and the shape of the incident laser pulse (plane wave versus Gaussian profile) on the acceleration rate. The results reveal a complex interplay between these parameters. For an electron beam with an initial energy of 10 MeV, a maximum accelerating rate of 345 MeV/m was achieved for the structure with a left-handed reflective grating and a base angle of $\alpha = 10^\circ$ when excited by a Gaussian pulse. Under plane-wave excitation, the left-handed reflective grating with $\alpha = 25^\circ$ provided a rate of 325 MeV/m. It is shown that for single-bunch injection, nearly the maximum energy gain can be attained within a temporal window of 0.5 fs. The obtained results provide a quantitative basis for optimizing DLA designs with triangular gratings, highlighting the significant role of the saw-tooth profile base angle and the grating type in achieving high rates for next-generation accelerators.
\end{abstract}

\maketitle

\section{\label{sec:intro}INTRODUCTION}

The development of compact, affordable sources of radiation and next-generation particle accelerators for research laboratories, medical, and industrial applications requires a paradigm shift compared to traditional microwave systems. Dielectric laser accelerators (DLAs) have become one of the leading candidates for addressing this challenge \cite{England2014,Wootton2016,Sotnikov2025,Andonian2025,BarLev2014, Simakov2017,Vasyliev2022,Bolshov2021,Vasyliev2023}. Experimental work on electron acceleration in dielectric micro- and nanostructures, as well as subsequent studies of the interaction of laser fields with subrelativistic electron beams, have demonstrated the potential of this approach for creating compact accelerating systems \cite{Oudheusden2007JAP, Rohloff2011, Kozak2018JAP}. One of the key directions for further research is the optimization of the geometry of dielectric structures and their laser excitation parameters, which makes it possible to increase the accelerating rate and improve the efficiency of energy transfer from the laser field to the electrons \cite{Xiria2015JAP, Chen2018}. By using intense ultrashort pulses from solid-state lasers to excite localized near fields within and around photonic structures, DLAs can generate accelerating fields exceeding 10 GV/m~\cite{Cesar2018}. This, combined with the low cost and scalability of fabrication technologies, makes them a promising foundation for the development of broadly accessible accelerator technologies \cite{Mourou2013}.

The operating principle of DLAs is based on the reversed Smith–Purcell effect \cite{Smith1953,Palmer1980}, in which a periodic dielectric structure synchronizes the laser-induced evanescent field with the electron beam. The geometry of the structure is of paramount importance, as it determines the spatial profile of the accelerating and decelerating field components. To date, most DLA research has focused on structures with rectangular or sinusoidal grating profiles \cite{McNeur2016, Leedle2015}. In contrast, DLAs with a triangular profile have not been studied in sufficient detail.

In Ref.\cite{Vasyliev2021}, a comparison of a chip-type triangular structure with other variants was carried out, but only one fixed geometry was considered. In a recent study \cite{Svystunov2025} triangular structures with different orientations and three fixed angles at the base of the "teeth" of the saw-tooth surface profile were investigated. The interest in studying triangular structures was driven, firstly, by the availability of commercially manufactured triangular structures with fixed base angles of the "teeth" (Thorlabs \cite{Thorlabs2024}, OpticElectronics), and, secondly, by the need to compare the acceleration rates in them with those in structures with a rectangular surface profile, which are traditionally studied in DLAs. It was shown that under certain conditions, triangular structures can provide a higher acceleration rate. A systematic analysis of the influence of the base angle ($\alpha$) over a wide range of values on the acceleration rate, especially in double triangular structures and in the presence of a reflective layer on one of the structures, is still lacking.

In this work, we fill this gap by conducting a comprehensive numerical study. Our goal is to quantitatively evaluate the influence of the angle ($\alpha$) at the base of the teeth of the accelerating grating on the achievable energy gain of electrons. We analyze a wide range of angles (from $5^\circ$ to $44^\circ$) for structures with double gratings, investigating the effect of orientation (left- and right-handed), and the presence or absence of a reflective coating on the second grating. Two cases of excitation of the accelerating field by a laser beam were studied: the plane-wave approximation and a Gaussian pulse. This systematic approach allows us to determine optimal geometric configurations, provide quantitative guidelines for future designs, and clarify the physical mechanisms managing electron acceleration in these structures. This study is part of a comprehensive effort on the development of dielectric laser accelerators being conducted at the National Science Center "Kharkiv Institute of Physics and Technology" \cite{Beznosenko2023}. For experimental studies on electron acceleration in DLAs, we have a femtosecond terawatt Ti:Sa laser system ~\cite{Vasiliev2018PAST}.

\section{\label{sec:methods}METHODS: PROBLEM FORMULATION AND SIMULATION}

Numerical simulations were performed using the Particle-In-Cell (PIC) method implemented in the CST Studio Suite software package \cite{CST}. This approach allows calculating the self-consistent interaction of the electron beam with the electromagnetic field excited by laser radiation.

\begin{figure}[!htb]
\includegraphics[width=\columnwidth]{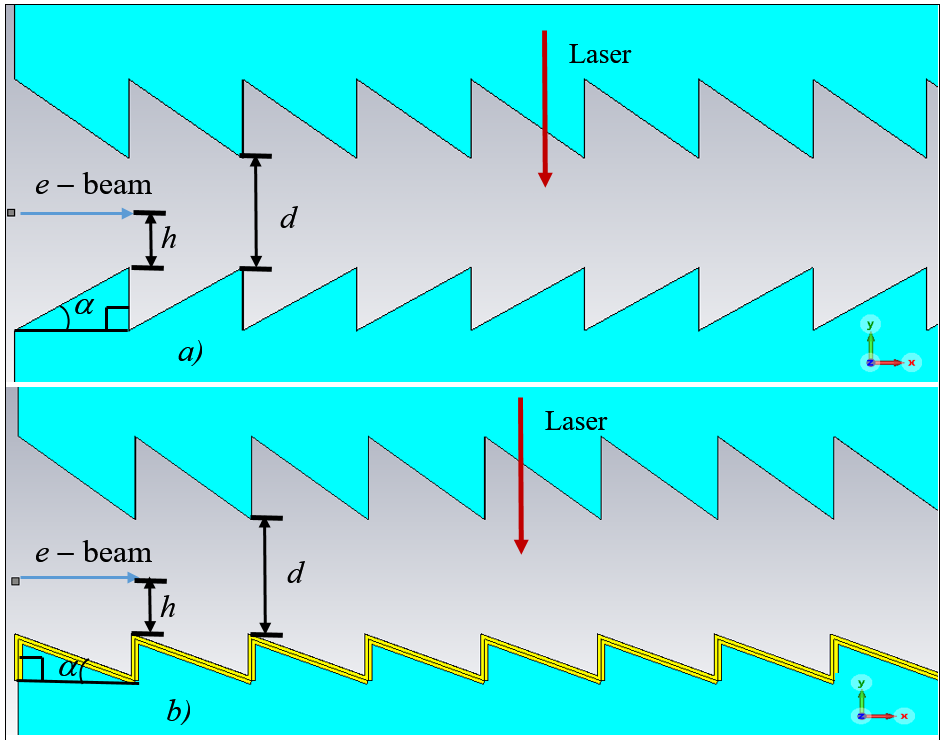}
\caption{\label{fig:fig1}Schematic of the model for numerical simulation of acceleration in a DLA with a triangular profile for double transparent-transparent (a) and
transparent-reflective (b) surfaces. For all investigated structures the upper transparent grating remained unchanged. The depiction of the lower gratings shows the right-handed and left-handed orientations of the triangular profile, respectively.}
\end{figure}

The geometry of the investigated structure with double gratings is shown in Fig.~\ref{fig:fig1}. The initial energy of the electron beam injected symmetrically into the structure along its X-axis (see Fig.~\ref{fig:fig1}) was 10 MeV. Such an electron beam energy is often used in numerical simulations and experimental studies of DLAs~\cite{Crisp2021,Lin2025,Sotnikov2025}.  The beam is formed by a cathode with a size of $50\times50$ nm. In the numerical calculations, it was modeled as a Direct Current (DC) beam with a very low current value, allowing space charge effects to be neglected. The DC model was chosen for systematic mapping of the accelerating potential of each structure, as it scans all phases of the field. Such a model provides a simple and clear identification of particles that have gained the maximum energy gain. The distance from the structure axis to the surface of the gratings is $h = 400$ nm. The study of acceleration of single short bunches synchronized in time with the injection of the laser Gaussian beam is presented in Section~\ref{sec:single}. The main simulation parameters are summarized in Table~\ref{tab:params}.

\begin{table}[!htb]
\caption{Simulation parameters of the DLA}\label{tab:params}
\begin{ruledtabular}
\begin{tabular}{|l|c|}
Parameter & Designation  / Value \\
\hline
Structure material & Fused silica \\
                   &($\varepsilon = 2.11$) \\
Grating period & $\lambda_p = 800$ nm \\
Laser wavelength & $\lambda_L = 800$ nm \\
Structure length & $L = 20\,\mu m$ \\
Gap between gratings & $d = 800$ nm \\
Base angle & \\
of upper grating (fixed) & $36^\circ$ \\
Base angle & \\
of lower grating (variable) & $\alpha = 5^\circ - 44^\circ$ \\
Conductivity of reflective coating  & $\sigma = 45.6\cdot 10^6$ S/m \\
 (gold) & \\
Initial electron energy & $W_{0} = 10$ MeV \\
Distance from beam axis  & $h = 400$ nm \\
to grating surface            & \\
Laser pulse field amplitude & $E_0 = 1$ GV/m \\
FWHM duration of Laser pulse      & $\tau_0=120$ fs\\
Diameter of Gaussian laser pulse waist        & $20\,\mu m$
\end{tabular}
\end{ruledtabular}
\end{table}

The laser pulse (a plane wave or a wave with a Gaussian profile) is incident perpendicularly onto the upper grating. In the simulations, the plane wave represented an idealized electromagnetic field of infinite transverse extent with a uniform distribution of amplitude and phase along the structure. The duration of the laser pulse with a Gaussian profile was 120 fs, and its waist diameter was equal to the length of the investigated structure. The field amplitude of the laser pulse for both the plane wave and the Gaussian profile was 1 GV/m. The total structure length was $L = 20\,\mu m$, and the grating period was $\lambda_p = 800$ nm, which approximately coincides with the laser wavelength $\lambda_L = 800$ nm, ensuring the fulfillment of the phase synchronization condition with the first spatial harmonic of the excited field\cite{Breuer2013,Palmer1980}:
\begin{equation}
\beta = \frac{v}{c} = \frac{\lambda_p}{\lambda_L} \frac{1}{n},
\label{eq:sync}
\end{equation}
where $n$ is the harmonic number, $\beta = v/c$ is the dimensionless electron velocity.

All investigated structures were double gratings. The upper grating with a fixed base angle of $36^\circ$ was transparent. The choice of this structure was based on the results of a preliminary numerical study, in which a comparative analysis of single transparent structures was performed while varying the base angle of the "tooth" of the triangular grating. The obtained results showed that the angle of $36^\circ$  provides the maximum energy gain of the electron beam. The parameters of the lower grating were varied depending on the configuration: a transparent right-handed grating (TR), a reflective right-handed grating (RR), and a reflective left-handed grating (RL). The material of the reflective coating was gold with a conductivity of $\sigma = 45.6\cdot 10^6$ S/m. Replacing the actual conductivity of gold with a lossless (perfect) conductor did not significantly change the results of the calculations regarding the energy gain of the accelerated electrons.

\section{\label{sec:results}RESULTS AND DISCUSSION}

\subsection{Plane-Wave Excitation}\label{sec:plane}

\begin{figure}[!htb]
\includegraphics[width=\columnwidth]{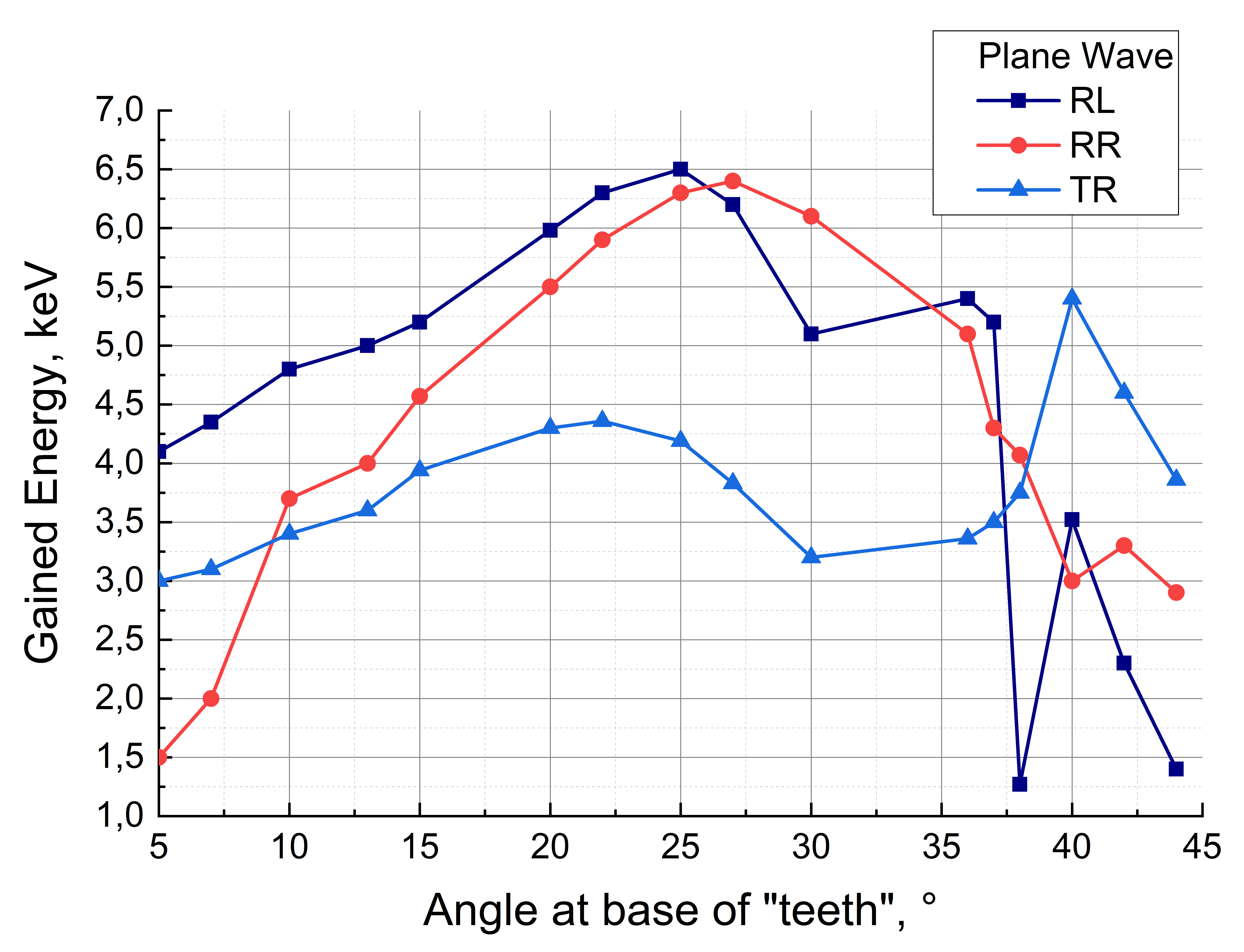}
\caption{\label{fig:fig2}Dependence of the energy gain of the maximally accelerated electrons of the beam on the base angle of the "tooth" of the lower grating in DLAs with double structures under plane-wave excitation. Here, TR is the transparent right-handed grating, RR is the reflective right-handed grating, and RL is the reflective left-handed grating. }
\end{figure}

The dependence of the energy gain of the maximally accelerated electrons of the beam on the base angle of the "tooth" of the lower grating ($\alpha$) for the case of plane-wave excitation is shown in Fig.~\ref{fig:fig2}. The maximum values are summarized in Table~\ref{tab:plane}. The obtained results are characterized by different dependencies for each configuration.

For the RR structure, the energy gain of the maximally accelerated particles increases monotonically with increasing $\alpha$ starting from small angle values, reaching a maximum of 6.4 keV at $\alpha = 27^\circ$ (accelerating rate of $320\, \rm MeV/m$). A broad region of high acceleration rates is observed in the range of $\alpha = 22^\circ$--$30^\circ$. The RL structure provides an absolute maximum of 6.5 keV (accelerating rate of $325\, \rm MeV/m$), but this peak is narrower and is achieved at $\alpha = 25^\circ$. The TR structure demonstrates a smoother, although lower, energy gain over the structure length, reaching a maximum of 5.4 keV at $\alpha = 40^\circ$ (accelerating rate of $270\, \rm MeV/m$).

These data show that to obtain high acceleration rates with lower tolerance requirements on the grating tooth angle, the RR structure is preferable. To achieve the absolute maximum energy gain, it is better to use the RL structure, but this requires higher manufacturing precision. The transparent triangular accelerating structure TR, although inferior in peak values, is an acceptable alternative when the use of a reflective coating is difficult or undesirable.

\begin{table}[!htb]
\caption{\label{tab:plane}Maximum energy gain $\Delta E$ (keV) of the beam electrons and accelerating rate $G = \Delta E/L$ (MeV/m) under excitation of the accelerating structure by a plane wave.}
\begin{ruledtabular}
\begin{tabular}{c|c|c|c|c|c|c}
 & \multicolumn{2}{c|}{RL} & \multicolumn{2}{c|}{RR} & \multicolumn{2}{c}{TR} \\
 \hline
$\alpha$, $^\circ$ & $\Delta E$, & $G$,  & $\Delta E$, & $G$, & $\Delta E$, & $G$, \\
 &  keV &  MeV/m &  keV & MeV/m &  keV &  MeV/m \\
\hline
5  & 4.1  & 205  & 1.5  & 75   & 3.0  & 150 \\
7  & 4.35 & 217.5 & 2.0  & 100  & 3.1  & 155 \\
10 & 4.8  & 240  & 3.7  & 185  & 3.4  & 170 \\
13 & 5.0  & 250  & 4.0  & 200  & 3.6  & 180 \\
15 & 5.2  & 260  & 4.57 & 228.5 & 3.94 & 197 \\
20 & 5.98 & 299  & 5.5  & 275  & 4.3  & 215 \\
22 & 6.3  & 315  & 5.9  & 295  & 4.36 & 218 \\
25 & \textbf{6.5} & \textbf{325} & 6.3  & 315  & 4.19 & 209.5 \\
27 & 6.2  & 310  & \textbf{6.4} & \textbf{320} & 3.83 & 191.5 \\
30 & 5.1  & 255  & 6.1  & 305  & 3.2  & 160 \\
36 & 5.4  & 270  & 5.1  & 255  & 3.36 & 168 \\
38 & 1.27 & 63.5 & 4.07 & 203.5 & 3.75 & 187.5 \\
40 & 3.52 & 176  & 3.0  & 150  & \textbf{5.4} & \textbf{270} \\
42 & 2.3  & 115  & 3.3  & 165  & 4.6  & 230 \\
44 & 1.4  & 70   & 2.9  & 145  & 3.86 & 193 \\
\end{tabular}
\end{ruledtabular}
\end{table}

\subsection{Excitation by a Wave with a Gaussian Profile}\label{sec:gauss}

\begin{figure}[!htb]
\includegraphics[width=\columnwidth]{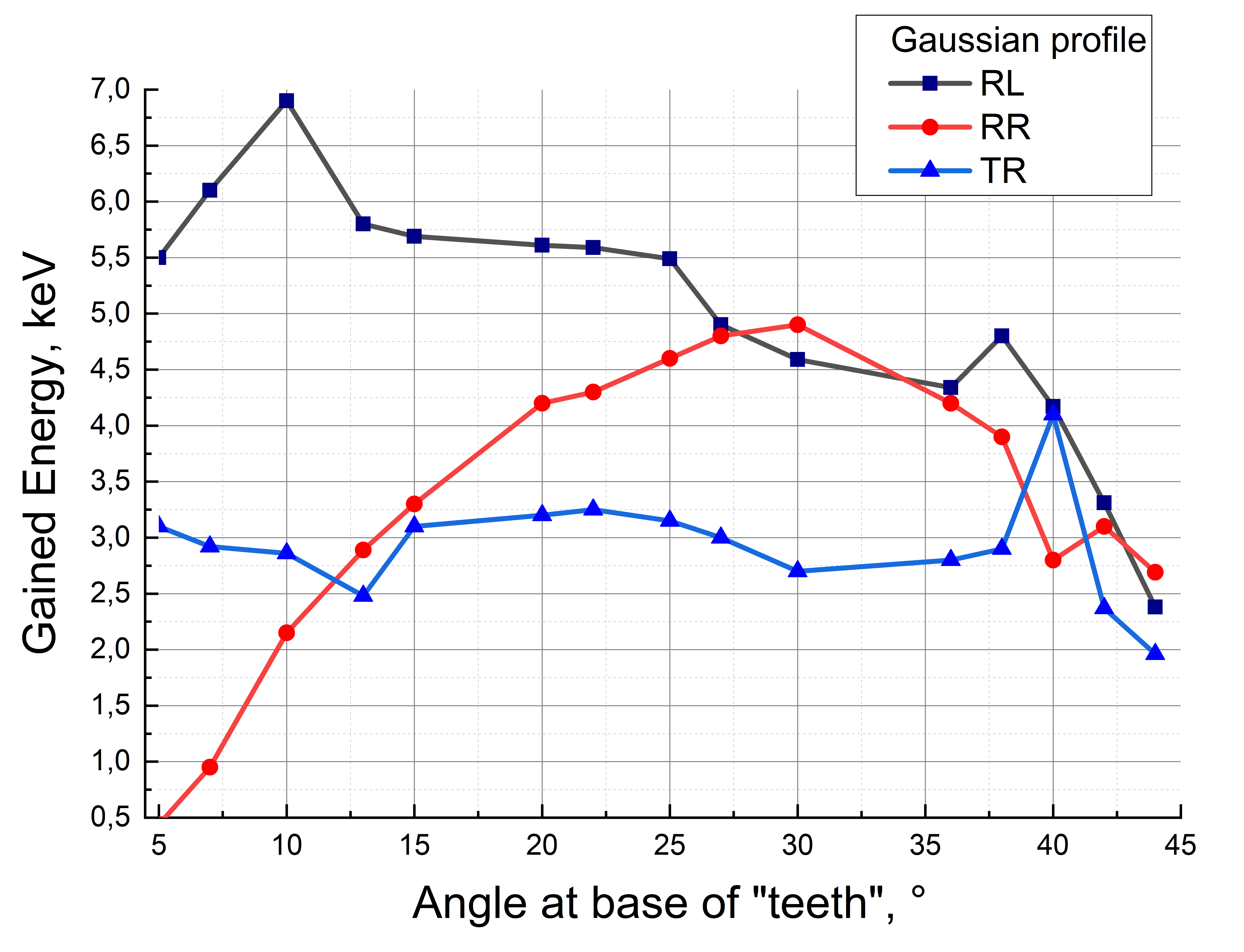}
\caption{\label{fig:fig3}Dependence of the energy gain of the maximally accelerated electrons of the beam on the base angle of the "tooth" of the lower grating in DLAs with double structures under excitation by a laser pulse with a Gaussian profile of the electromagnetic field amplitude distribution. Here, TR is the transparent right-handed grating, RR is the reflective right-handed grating, and RL is the reflective left-handed grating.}
\end{figure}

Realistic laser pulses have a Gaussian spatial and temporal profile with a finite duration, which significantly affects the spatial shape of the excited field inside the channel of the electron beam passage. In the results presented below, a pulse with an FWHM duration of 120 fs and a waist diameter of $20\,\mu m$ was used, and the electric field amplitude was $E_0=1$ GV/m. The calculation results are shown in Fig.~\ref{fig:fig3} and Table~\ref{tab:gauss}.

The most significant difference is observed for the RL structure, where the optimal angle shifts from $\alpha = 25^\circ$ (for the plane wave) to a wide range of $\alpha = 5^\circ - 15^\circ$ with a peak of 6.9 keV at $\alpha = 10^\circ$. This may be due to more efficient matching of the spatial distribution of the amplitude and phase profile of the Gaussian beam with the geometry of the structure, which contributes to an increase in the accelerating field in the electron beam passage channel and, consequently, to an increase in the acceleration rate. The RR structure demonstrates close values of the acceleration rate over a wide range of angles $\alpha = 25^\circ-37^\circ$ with an energy gain of $4.5-5.0$ keV (acceleration rate of $225-250\,\rm MeV/m$). The TR structure reaches a maximum of 4.1 keV at $\alpha = 40^\circ$ (acceleration rate $205\, \rm MeV/m$), repeating the trend observed in the numerical simulations under the plane-wave approximation.

The shift of the optimal $\alpha$, obtained for the RL and RR structures is a key result. It shows that although the plane-wave approximation allows rapid qualitative results of the electron acceleration process to be obtained, it is insufficient for accurate prediction of the characteristics of DLAs with real laser pulse profiles. The reduction in energy gain in all configurations is explained by the non-uniform amplitude, wavefront curvature, and finite transverse size of the Gaussian beam.

\begin{table}[!htb]
\caption{\label{tab:gauss}Maximum energy gain $\Delta E$ (keV) of the beam electrons and accelerating rate $G = \Delta E/L$ (MeV/m) under excitation by a pulse with a Gaussian profile of the electromagnetic field amplitude distribution.}
\begin{ruledtabular}
\begin{tabular}{c|c|c|c|c|c|c}
 & \multicolumn{2}{c|}{RL} & \multicolumn{2}{c|}{RR} & \multicolumn{2}{c}{TR} \\
 \hline
$\alpha$, $^\circ$ & $\Delta E$, & $G$,  & $\Delta E$, & $G$, & $\Delta E$, & $G$, \\
 &  keV &  MeV/m &  keV & MeV/m &  keV &  MeV/m \\
\hline
5  & 5.5  & 275   & 0.44 & 22   & 3.1  & 155 \\
7  & 6.1  & 305   & 0.95 & 47.5 & 2.92 & 146 \\
10 & \textbf{6.9} & \textbf{345} & 2.15 & 107.5 & 2.86 & 143 \\
13 & 5.8  & 290   & 3.3  & 165  & 2.48 & 124 \\
15 & 5.69 & 284.5 & 3.3  & 165  & 3.1  & 155 \\
20 & 5.61 & 280.5 & 4.2  & 210  & 3.2  & 160 \\
22 & 5.59 & 279.5 & 4.3  & 215  & 3.25 & 162.5 \\
25 & 5.49 & 274.5 & 4.6  & 230  & 3.15 & 157.5 \\
27 & 4.9  & 245   & 4.8  & 240  & 3.0  & 150 \\
30 & 4.59 & 229.5 & \textbf{4.9} & \textbf{245} & 2.7  & 135 \\
36 & 4.34 & 217   & 4.2  & 210  & 2.8  & 140 \\
38 & 4.8  & 240   & 3.9  & 195  & 2.9  & 145 \\
40 & 4.17 & 208.5 & 2.8  & 140  & \textbf{4.1} & \textbf{205} \\
42 & 3.31 & 165.5 & 3.1  & 155  & 2.37 & 118.5 \\
44 & 2.38 & 119   & 2.69 & 134.5 & 1.96 & 98 \\
\end{tabular}
\end{ruledtabular}
\end{table}

\subsection{Acceleration of Single Electrons and Short Bunches}\label{sec:single}

\begin{figure}[!tbh]
\includegraphics[width=\columnwidth]{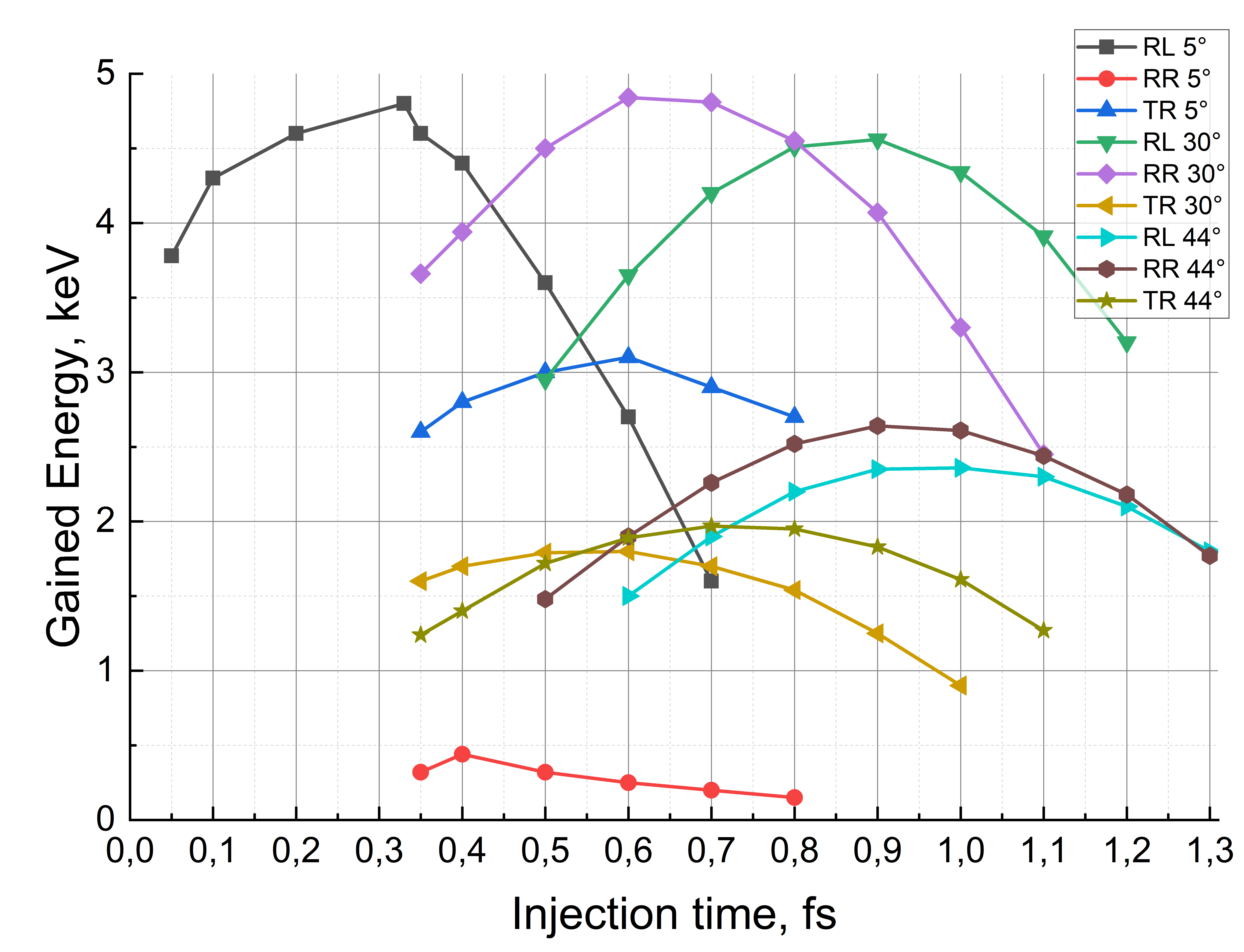}
\caption{\label{fig:fig4}Dependence of the electron energy gain on the injection time for various periodic accelerating structures. The curve labels TR, RR, and RL are the same as in Fig.~\ref{fig:fig2} and Fig.~\ref{fig:fig3}.}
\end{figure}

The dependences presented in Fig.~\ref{fig:fig2} and Fig.~\ref{fig:fig3} show the energy of the maximally accelerated electron at the end of the DLA section. However, this may be an electron that was initially in a decelerating phase of the field, since we used DC injection of the electron beam, in which the particles are uniformly distributed over the phase of the accelerating field. It is of practical interest to investigate the dynamics of a short electron bunch, the length of which is significantly smaller than the central wavelength of the laser pulse.. For this purpose, the acceleration of a point-like electron bunch was studied. The optimal injection time (allowing the maximum energy gain to be obtained) relative to the laser pulse with a Gaussian profile was preliminarily determined. The pulse duration was the same as for the results presented in Fig.~\ref{fig:fig3}, i.e., 120 fs. This synchronization was performed for the base angle values of the "tooth" $\alpha = 5^\circ$, $30^\circ$ and $44^\circ$.  The values of $\alpha = 5^\circ$ and $44^\circ$ correspond to the lower and upper limits of the investigated angle range, while the value of  $\alpha = 30^\circ$ represents an intermediate case, allowing the features of single-electron acceleration to be evaluated for a grating geometry that does not correspond to the extreme configurations. The capture temporal window for the investigated structures is shown in Fig.~\ref{fig:fig4}. The simulation shows that all structures have a capture window of about 0.5 fs. However, to achieve the maximum energy gain, injection must be performed with an accuracy of about 0.1 fs. This underscores the need for precise synchronization of the electron bunch and the laser pulse. The slight shift of the optimal injection time for the RL and RR structures is explained by the different phase velocities of the excited surface waves, which is a direct consequence of their geometry.

It should be noted that initially all dependences in Fig.~\ref{fig:fig4} were calculated starting from the laser pulse injection time $t = 0$ fs. For the RL $5^\circ$ curve, the initial calculation range turned out to be insufficient to display the optimal injection time. Therefore, for this structure, the simulation was continued up to 2.3 fs. In this case, after the laser pulse passed through part of the structure, the field configuration repeated the initial state, which made it possible to reconstruct the missing portion of the dependence and add three points to the left of the maximum. For the other dependences, a similar correction of the time scale was performed: the curves were shifted by 0.3 fs to ensure their comparability with the reconstructed RL $5^\circ$ dependence.

\begin{figure}[!htb]
\includegraphics[width=\columnwidth]{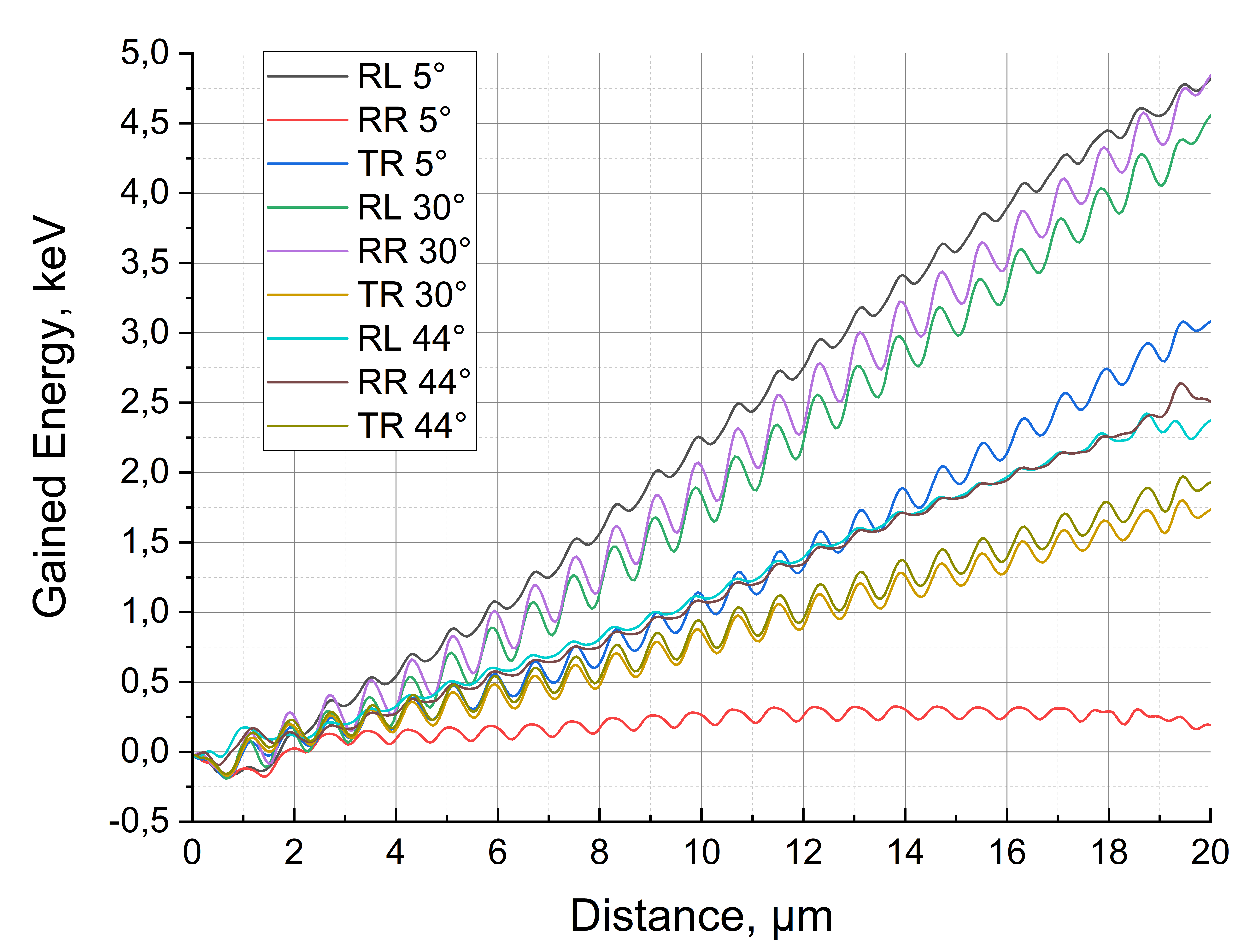}
\caption{\label{fig:fig5}Change in the electron energy gain along the periodic structures at the optimal injection time. The curve labels TR, RR, and RL are the same as in Fig~\ref{fig:fig2} and Fig.~\ref{fig:fig3}.}
\end{figure}

Fig.~\ref{fig:fig5} shows the energy gain change along the structure at the optimal delay. It should be noted that the RL structure with $\alpha = 5^\circ$,  $30^\circ$ and the RR structure with $\alpha =  30^\circ$ provide the most stable energy gain with minimal influence of the decelerating phase.

The small difference between the energy gain of the maximally accelerated particles (Fig.~\ref{fig:fig2}, Fig.~\ref{fig:fig3}) and the energy gain of a single electron injected with the optimal delay time (Fig.~\ref{fig:fig4}) is explained as follows.
In the study presented in Fig.~\ref{fig:fig4} and ~\ref{fig:fig5}, the electron was injected exactly from the geometric center of the emitting surface of the electron gun. In contrast, for the data in Fig.~\ref{fig:fig2} and ~\ref{fig:fig3} electron emission from the entire cathode area was simulated, which resulted in a beam with a wide spread of initial electron coordinates. In the latter case, some electrons propagate at a smaller height above the structures, where the field gradient is higher than along the axis of the geometric center of the cathode. This explains the observation of higher energy values.

\section{\label{sec:conclusion}CONCLUSION}

In this work, a numerical study of electron acceleration in dielectric laser accelerators based on double gratings with a triangular profile has been performed. Using PIC simulations, the influence of the base angle of the lower grating ($\alpha = 5^\circ$--$44^\circ$), its orientation, the presence of a reflective coating, and the shape of the laser pulse on the electron energy gain was systematically investigated. The main results are:
\begin{enumerate}
  \item The base angle is a critical parameter. Its optimal value strongly depends both on the structure configuration and on the laser pulse shape. For the RL structure (reflective grating with leftward-sloping "teeth"), the optimal angle is $\alpha \approx 25^\circ$ under excitation in the plane-wave approximation. Under the Gaussian laser pulse profile, the optimal angle is $\alpha \approx 10^\circ$.
  \item Reflective coatings on one of the double gratings significantly increase the acceleration rate. Over a DLA length of $20\,\mu m$ , the RL structure showed the highest energy gain: 6.9 keV (rate of 345 MeV/m) for a Gaussian beam at $\alpha = 10^\circ$ and 6.5 keV (325 MeV/m) for a plane wave at $\alpha = 25^\circ$.
  \item The plane-wave approximation can be used for a fast qualitative evaluation of the acceleration rates in DLAs with double triangular structures. However, for quantitative calculations of electron acceleration characteristics in DLAs, the results of which can be used for verification in experimental studies, simulations using a Gaussian laser pulse profile are necessary. The shift in optimal $\alpha$ and the overall reduction in energy gain compared to the plane-wave case emphasize the need to use realistic laser pulse models.
  \item Acceleration of a single bunch is possible within a narrow but physically realizable temporal window. A capture window of about 0.5 fs was found for all structures, which requires high-precision synchronization of electron bunches with the laser pulse.
\end{enumerate}

The obtained results provide a quantitative framework for the optimization of DLAs with triangular gratings. The conducted studies identify the base angle as a key parameter for achieving high accelerating rates, with the reflector geometry being an additional key parameter specific to reflective structures, and provide practical guidelines for the development of next-generation accelerators based on gratings with a triangular surface profile. This study is part of a systematic effort on the development of dielectric laser accelerators being carried out at the National Science Center "Kharkiv Institute of Physics and Technology" \cite{Beznosenko2023}.

\begin{acknowledgments}
The study is supported by the National Research Foundation of Ukraine under the program “Excellent Science in Ukraine” (project No. 2023.03/0182).
\end{acknowledgments}

\section*{AUTHOR DECLARATIONS}

\subsection*{Conflict of Interest}
The authors declare no conflict of interest.

\subsection*{Author Contributions}
\textbf{O.O. Svystunov:} Conceptualization (equal); Data curation (lead); Formal analysis (equal); Investigation (lead); Methodology (equal); Visualization (lead); Writing – original draft (lead); Writing – review and editing (equal). \\
\textbf{A.V. Vasyliev:} Conceptualization (equal); Formal analysis (supporting); Supervision (equal); Writing – review and editing (equal). \\
\textbf{I.V. Beznosenko:} Investigation (supporting); Methodology (supporting); Writing – review and editing (supporting). \\
\textbf{R.R. Knyazev:} Methodology (supporting); Writing – review and editing (supporting).\\
\textbf{G.V. Sotnikov:} Conceptualization (lead); Formal analysis (equal); Funding acquisition (lead); Project administration (lead); Supervision (lead); Writing – review and editing (equal).

\section*{DATA AVAILABILITY STATEMENT}

The data that supports the findings of this study are available
within the article. Additional data supporting the findings of this study are available from the corresponding author upon reasonable request.

\bibliography{Svystunov}

\end{document}